\documentstyle[epsfig,a4p,12pt,rotating]{article}
\begin{document}
%
%
\newcommand{\PPEnum}    {CERN-EP/2003-016}
\newcommand{\PNnum}     {OPAL Physics Note PN-497}
\newcommand{\TNnum}     {OPAL Technical Note TN-xxx}
\newcommand{\Date}      {11 April 2003}
\newcommand{\Author}    {G.~Benelli, S.~Braibant, M.~Fanti, P.~Giacomelli and P.~Tran}
\newcommand{\MailAddr}  {paolo.giacomelli@cern.ch}
\newcommand{\EdBoard}   {C.~Chang, P.~Gagnon, M.~Hauschild and J.~Pinfold}
\newcommand{\DraftVer}  {Version 3.0}
\newcommand{\DraftDate} {\today}
\newcommand{\TimeLimit} {June 27, 14:00}

\def\toprule{\noalign{\hrule \medskip}}
\def\midrule{\noalign{\medskip\hrule }}
\def\botrule{\noalign{\medskip\hrule }}

\setlength{\parskip}{\medskipamount}


\newcommand{\LL}{{\mathrm L}^+ {\mathrm L}^-}
\newcommand{\QQ}{{\mathrm Q}\bar{\mathrm Q}}
\newcommand{\XX}{{\mathrm X}^+ {\mathrm X}^-}
\newcommand{\ee}{{\mathrm e}^+ {\mathrm e}^-}
\newcommand{\sq}{\tilde{\mathrm q}}
\newcommand{\seff}{\tilde{\mathrm f}}
\newcommand{\sele}{\tilde{\mathrm e}}
\newcommand{\sell}{\tilde{\ell}}
\newcommand{\snu}{\tilde{\nu}}
\newcommand{\supq}{\tilde{u}}
\newcommand{\sdown}{\tilde{d}}
\newcommand{\smu}{\tilde{\mu}}
\newcommand{\stau}{\tilde{\tau}}
\newcommand{\chp}{\tilde{\chi}^+_1}
\newcommand{\chn}{\tilde{\chi}^+_n}
\newcommand{\chpm}{\tilde{\chi}^\pm_1}
\newcommand{\nt}{\tilde{\chi}^0}
\newcommand{\qq}{{\mathrm q}\bar{\mathrm q}}
\newcommand{\sleppair}{\sell^+ \sell^-}
\newcommand{\nunu}{\nu \bar{\nu}}
\newcommand{\mumu}{\mu^+ \mu^-}
\newcommand{\tautau}{\tau^+ \tau^-}
\newcommand{\ellell}{\ell^+ \ell^-}
\newcommand{\nulqq}{\nu \ell {\mathrm q} \bar{\mathrm q}'}
\newcommand{\MZ}{M_{\mathrm Z}}
\newcommand{\eeff}{\ee\mathrm{f\bar{f}}}
\newcommand{\msl}{$m_{\sell}$}

\newcommand {\stopm}         {\tilde{\mathrm{t}}_{1}}
\newcommand {\stops}         {\tilde{\mathrm{t}}_{2}}
\newcommand {\stopbar}       {\bar{\tilde{\mathrm{t}}}_{1}}
\newcommand {\stopx}         {\tilde{\mathrm{t}}}
\newcommand {\sneutrino}     {\tilde{\nu}}
\newcommand {\slepton}       {\tilde{\ell}}
\newcommand {\stopl}         {\tilde{\mathrm{t}}_{\mathrm L}}
\newcommand {\stopr}         {\tilde{\mathrm{t}}_{\mathrm R}}
\newcommand {\stoppair}      {\tilde{\mathrm{t}}_{1}
\bar{\tilde{\mathrm{t}}}_{1}}
\newcommand {\gluino}        {\tilde{\mathrm g}}

\newcommand {\neutralino}    {\tilde{\chi }^{0}_{1}}
\newcommand {\neutrala}      {\tilde{\chi }^{0}_{2}}
\newcommand {\neutralb}      {\tilde{\chi }^{0}_{3}}
\newcommand {\neutralc}      {\tilde{\chi }^{0}_{4}}
\newcommand {\bino}          {\tilde{\mathrm B}^{0}}
\newcommand {\wino}          {\tilde{\mathrm W}^{0}}
\newcommand {\higginoa}      {\tilde{\rm H_{1}}^{0}}
\newcommand {\higginob}      {\tilde{\mathrm H_{1}}^{0}}
\newcommand {\chargino}      {\tilde{\chi }^{\pm}_{1}}
\newcommand {\chargib}       {\tilde{\chi }^{\pm}_{2}}
\newcommand {\charginop}     {\tilde{\chi }^{+}_{1}}
\newcommand {\chargibp}      {\tilde{\chi }^{+}_{2}}
\newcommand {\KK}            {{\mathrm K}^{0}-\bar{\mathrm K}^{0}}
\newcommand {\ff}            {{\mathrm f} \bar{\mathrm f}}
\newcommand {\bstopm} {\mbox{$\boldmath {\tilde{\mathrm{t}}_{1}} $}}
\newcommand {\Mt}            {M_{\mathrm t}}
\newcommand {\mscalar}       {m_{0}}
\newcommand {\Mgaugino}      {M_{1/2}}
\newcommand {\rs}            {\sqrt{s}}
\newcommand {\WW}            {{\mathrm W}^+{\mathrm W}^-}
\newcommand {\MGUT}          {M_{\mathrm {GUT}}}
\newcommand {\Zboson}        {{\mathrm Z}^{0}}
\newcommand {\Wpm}           {{\mathrm W}^{\pm}}
\newcommand {\allqq}         {\sum_{q \neq t} q \bar{q}}
\newcommand {\mixang}        {\theta _{\mathrm {mix}}}
\newcommand {\thacop}        {\theta _{\mathrm {Acop}}}
\newcommand {\cosjet}        {\cos\thejet}
\newcommand {\costhr}        {\cos\thethr}
\newcommand {\djoin}         {d_{\mathrm{join}}}
\newcommand {\mstop}         {m_{\stopm}}
\newcommand {\msell}         {m_{\sell}}
\newcommand {\mchi}          {m_{\neutralino}}
\newcommand {\pp}{p \bar{p}}

\newcommand{\epair}{\mbox{${\mathrm e}^+{\mathrm e}^-$}}
\newcommand{\mupair}{\mbox{$\mu^+\mu^-$}}
\newcommand{\taupair}{\mbox{$\tau^+\tau^-$}}
\newcommand{\qpair}{\mbox{${\mathrm q}\overline{\mathrm q}$}}
\newcommand{\eeee}{\mbox{\epair\epair}}
\newcommand{\eemumu}{\mbox{\epair\mupair}}
\newcommand{\eetautau}{\mbox{\epair\taupair}}
\newcommand{\eeqq}{\mbox{\epair\qpair}}
\newcommand{\fs}{ final states}
\newcommand{\epairf}{\mbox{\epair\fs}}
\newcommand{\mupairf}{\mbox{\mupair\fs}}
\newcommand{\taupairf}{\mbox{\taupair\fs}}
\newcommand{\qpairf}{\mbox{\qpair\fs}}
\newcommand{\eeeef}{\mbox{\eeee\fs}}
\newcommand{\eemumuf}{\mbox{\eemumu\fs}}
\newcommand{\eetautauf}{\mbox{\eetautau\fs}}
\newcommand{\eeqqf}{\mbox{\eeqq\fs}}
\newcommand{\ffff}{four fermion final states}
\newcommand{\llnunu}{\mbox{\lpair\nul\nubar}}
\newcommand{\lnuqq}{\mbox{\lept\nubar\qpair}}
\newcommand{\zee}{\mbox{Zee}}
\newcommand{\zzg}{\mbox{ZZ/Z$\gamma$}}
\newcommand{\wenu}{\mbox{We$\nu$}}

\newcommand{\el}{\mbox{${\mathrm e}^-$}}
\newcommand{\selem}{\mbox{$\tilde{\mathrm e}^-$}}
\newcommand{\smum}{\mbox{$\tilde\mu^-$}}
\newcommand{\staum}{\mbox{$\tilde\tau^-$}}
\newcommand{\slept}{\mbox{$\tilde{\ell}^\pm$}}
\newcommand{\sleptm}{\mbox{$\tilde{\ell}^-$}}
\newcommand{\lept}{\mbox{$\ell^-$}}
\newcommand{\Hl}{\mbox{$\mathrm{L}^\pm$}}
\newcommand{\Hm}{\mbox{$\mathrm{L}^-$}}
\newcommand{\Hnu}{\mbox{$\nu_{\mathrm{L}}$}}
\newcommand{\nul}{\mbox{$\nu_\ell$}}
\newcommand{\nubar}{\mbox{$\overline{\nu}_\ell$}}
\newcommand{\spair}{\mbox{$\tilde{\ell}^+\tilde{\ell}^-$}}
\newcommand{\lpair}{\mbox{$\ell^+\ell^-$}}
\newcommand{\staupair}{\mbox{$\tilde{\tau}^+\tilde{\tau}^-$}}
\newcommand{\smupair}{\mbox{$\tilde{\mu}^+\tilde{\mu}^-$}}
\newcommand{\selepair}{\mbox{$\tilde{\mathrm e}^+\tilde{\mathrm e}^-$}}
\newcommand{\ch}{\mbox{$\tilde{\chi}^\pm_1$}}
\newcommand{\chpair}{\mbox{$\tilde{\chi}^+_1\tilde{\chi}^-_1$}}
\newcommand{\chm}{\mbox{$\tilde{\chi}^-_1$}}
\newcommand{\chmp}{\mbox{$\tilde{\chi}^\pm_1$}}
\newcommand{\chz}{\mbox{$\tilde{\chi}^0_1$}}
\newcommand{\dch}{\mbox{\chm$\rightarrow$\chz\lept\nubar}}
\newcommand{\dslept}{\mbox{\sleptm$\rightarrow$\chz\lept}}
\newcommand{\dH}{\mbox{\Hm$\rightarrow$\lept\nubar\Hnu}}
\newcommand{\mch}{\mbox{$m_{\tilde{\chi}^\pm_1}$}}
\newcommand{\mslept}{\mbox{$m_{\tilde{\ell}}$}}
\newcommand{\mstau}{\mbox{$m_{\staum}$}}
\newcommand{\msmu}{\mbox{$m_{\smum}$}}
\newcommand{\msele}{\mbox{$m_{\selem}$}}
\newcommand{\mchz}{\mbox{$m_{\tilde{\chi}^0_1}$}}
\newcommand{\dm}{\mbox{$\Delta m$}}
\newcommand{\dmch}{\mbox{$\Delta m_{\ch-\chz}$}}
\newcommand{\dmslept}{\mbox{$\Delta m_{\slept-\chz}$}}
\newcommand{\dmhl}{\mbox{$\Delta m_{\Hl-\Hnu}$}}
\newcommand{\w}{\mbox{W$^\pm$}}

\newcommand{\acopc}{\mbox{$\phi^{\mathrm{acop}}$}}
\newcommand{\acolc}{\mbox{$\theta^{\mathrm{acol}}$}}
\newcommand{\acop}{\mbox{$\phi_{\mathrm{acop}}$}}
\newcommand{\acol}{\mbox{$\theta_{\mathrm{acol}}$}}
\newcommand{\pt}{\mbox{$p_{t}$}}
\newcommand{\pz}{\mbox{$p_{\mathrm{z}}^{\mathrm{miss}}$}}
\newcommand{\ptevt}{\mbox{$p_{t}^{\mathrm{miss}}$}}
\newcommand{\ptaxic}{\mbox{$a_{t}^{\mathrm{miss}}$}}
\newcommand{\stevt}{\mbox{$p_{t}^{\mathrm{miss}}$/\Ebeam}}
\newcommand{\staxic}{\mbox{$a_{t}^{\mathrm{miss}}$/\Ebeam}}
\newcommand{\dptaxic}{\mbox{missing $p_{t}$ wrt. event axis \ptaxic}}
\newcommand{\cosevt}{\mbox{$\mid\cos\theta_{\mathrm{p}}^{\mathrm{miss}}\mid$}}
\newcommand{\axicos}{\mbox{$\mid\cos\theta_{\mathrm{a}}^{\mathrm{miss}}\mid$}}
\newcommand{\pthet}{\mbox{$\theta_{\mathrm{p}}^{\mathrm{miss}}$}}
\newcommand{\athet}{\mbox{$\theta_{\mathrm{a}}^{\mathrm{miss}}$}}
\newcommand{\dcosevt}{\mbox{$\mid\cos\theta\mid$ of missing p$_{t}$}}
\newcommand{\daxicos}{\mbox{$\mid\cos\theta\mid$ of missing p$_{t}$ wrt. event
axis}}
\newcommand{\efdsw}{\mbox{$x_{\mathrm{FDSW}}$}}
\newcommand{\acopf}{\mbox{$\Delta\phi_{\mathrm{FDSW}}$}}
\newcommand{\acopm}{\mbox{$\Delta\phi_{\mathrm{MUON}}$}}
\newcommand{\acopt}{\mbox{$\Delta\phi_{\mathrm{trk}}$}}
\newcommand{\po}{\mbox{$E_{\mathrm{isol}}^\gamma$}}
\newcommand{\qprod}{\mbox{$q1$$*$$q2$}}
\newcommand{\lcode}{lepton identification code}
\newcommand{\nctro}{\mbox{$N_{\mathrm{trk}}^{\mathrm{out}}$}}
\newcommand{\necao}{\mbox{$N_{\mathrm{ecal}}^{\mathrm{out}}$}}
\newcommand{\mout}{\mbox{$m^{\mathrm{out}}$}}
\newcommand{\nctec}{\mbox{\nctro$+$\necao}}
\newcommand{\gfract}{\mbox{$F_{\mathrm{good}}$}}
\newcommand{\zz}       {\mbox{$|z_0|$}}
\newcommand{\dz}       {\mbox{$|d_0|$}}
\newcommand{\sint}      {\mbox{$\sin\theta$}}
\newcommand{\cost}      {\mbox{$\cos\theta$}}
\newcommand{\mcost}     {\mbox{$|\cos\theta|$}}
\newcommand{\dedx}     {{\mathrm d}E/{\mathrm d}x}
\newcommand{\wdedx}     {\mbox{$W_{dE/dx}$}}
\newcommand{\xe}     {\mbox{$x_E$}}

\newcommand{\ssix}     {\mbox{$\sqrt{s}$~=~161~GeV}}
\newcommand{\sthree}     {\mbox{$\sqrt{s}$~=~130--136~GeV}}
\newcommand{\mrecoil}     {\mbox{$m_{\mathrm{recoil}}$}}
\newcommand{\llmass}     {\mbox{$m_{ll}$}}
\newcommand{\sml}{\mbox{Standard Model \lpair$\nu\nu$ events}}
\newcommand{\sme}{\mbox{Standard Model events}}
\newcommand{\sig}
  {events containing a lepton pair plus missing transverse momentum}
\newcommand{\wpair}{\mbox{$W^+W^-$}}
\newcommand{\dW}{\mbox{W$^-\rightarrow$\lept\nubar}}
\newcommand{\dsele}{\mbox{\selem$\rightarrow$\chz e$^-$}}
\newcommand{\eeeell}{\mbox{\epair$\rightarrow$\epair\lpair}}
\newcommand{\eell}{\mbox{\epair\lpair}}
\newcommand{\llgam}{\mbox{$\ell\ell(\gamma)$}}
\newcommand{\nunugam}{\mbox{$\nu\bar{\nu}\gamma\gamma$}}
\newcommand{\acope}{\mbox{$\Delta\phi_{\mathrm{EE}}$}}
\newcommand{\nee}{\mbox{N$_{\mathrm{EE}}$}}
\newcommand{\eesum}{\mbox{$\Sigma_{\mathrm{EE}}$}}
\newcommand{\at}{\mbox{$a_{t}$}}
\newcommand{\spp}{\mbox{$p$/\Ebeam}}
\newcommand{\acoph}{\mbox{$\Delta\phi_{\mathrm{HCAL}}$}}

\newcommand{\roots}     {\sqrt{s}}
%
%
\newcommand{\thrust}    {T}
\newcommand{\nthrust}   {\hat{n}_{\mathrm{thrust}}}
\newcommand{\thethr}    {\theta_{\,\mathrm{thrust}}}
\newcommand{\phithr}    {\phi_{\mathrm{thrust}}}
\newcommand{\acosthr}   {|\cos\thethr|}
\newcommand{\thejet}    {\theta_{\,\mathrm{jet}}}
\newcommand{\acosjet}   {|\cos\thejet|}
\newcommand{\thmiss}    { \theta_{miss} }
\newcommand{\cosmiss}   {| \cos \thmiss |}

\newcommand{\Evis}      {E_{\mathrm{vis}}}
\newcommand{\Rvis}      {E_{\mathrm{vis}}\,/\roots}
\newcommand{\Mvis}      {M_{\mathrm{vis}}}
\newcommand{\Rbal}      {R_{\mathrm{bal}}}

\newcommand{\Ecm}{\mbox{$E_{\mathrm{cm}}$}}
\newcommand{\Ebeam}{\mbox{$E_{\mathrm{beam}}$}}
\newcommand{\ipb}{\mbox{pb$^{-1}$}}
\newcommand{\wrt}{with respect to}
\newcommand{\sm}{Standard Model}
\newcommand{\smb}{Standard Model background}
\newcommand{\smp}{Standard Model processes}
\newcommand{\smc}{Standard Model Monte Carlo}
\newcommand{\mc}{Monte Carlo}
\newcommand{\btb}{back-to-back}
\newcommand{\tp}{two-photon}
\newcommand{\tpb}{two-photon background}
\newcommand{\tpp}{two-photon processes}
\newcommand{\lp}{lepton pairs}
\newcommand{\vto}{\mbox{$\tau$ veto}}
\newcommand{\gsim}{\;\raisebox{-0.9ex}
           {$\textstyle\stackrel{\textstyle >}{\sim}$}\;}
\newcommand{\lsim}{\;\raisebox{-0.9ex}{$\textstyle\stackrel{\textstyle<}
           {\sim}$}\;}
\newcommand{\degree}    {^\circ}
\newcommand{\Rparity}{$R$-parity}
\newcommand{\Rp}  {$R_{p}$}  
\newcommand{\lb}  {$\lambda$}
\newcommand{\lbp} {$\lambda^{'}$}
\newcommand{\lbpp} {$\lambda^{''}$}

\newcommand{\phiacop}   {\phi_{\mathrm{acop}}}


%
%
\newcommand{\ZP}[3]    {Z. Phys. {\bf C#1} (#2) #3.}
\newcommand{\PL}[3]    {Phys. Lett. {\bf B#1} (#2) #3.}
\newcommand{\etal}     {{\it et al}.,\,\ }
\newcommand{\PhysLett}  {Phys.~Lett.}
\newcommand{\PRL} {Phys.~Rev.\ Lett.}
\newcommand{\PhysRep}   {Phys.~Rep.}
\newcommand{\PhysRev}   {Phys.~Rev.}
\newcommand{\NPhys}  {Nucl.~Phys.}
\newcommand{\NIM} {Nucl.~Instr.\ Meth.}
\newcommand{\CPC} {Comp.~Phys.\ Comm.}
\newcommand{\ZPhys}  {Z.~Phys.}
\newcommand{\IEEENS} {IEEE Trans.\ Nucl.~Sci.}
%
%
\newcommand{\OPALColl}   {OPAL Collab.}
\newcommand{\DELPHIColl} {DELPHI Collab.}
\newcommand{\ALEPHColl}  {ALEPH Collab.}
\newcommand{\JADEColl}  {JADE Collab.}
%
\newcommand{\onecol}[2] {\multicolumn{1}{#1}{#2}}
\newcommand{\ra}        {\rightarrow}   

\newcommand {\qqp}           {\mbox{$\mathrm{q\overline{q}^\prime}$}}
\newcommand {\qpq}           {\mbox{$\mathrm{q^\prime\overline{q}}$}}
\newcommand {\Z}             {\mbox{${\mathrm Z}^{0}$}}
\newcommand {\W}             {\mbox{W$^{\pm}$}}


\begin{titlepage}
\begin{center}
    \large
    EUROPEAN ORGANIZATION FOR NUCLEAR RESEARCH 
\end{center}
\begin{flushright}
    \large
    \PPEnum\\
    \Date
\end{flushright}


\begin{center}
    \huge\bf\boldmath
 Search for Stable and Long-Lived Massive Charged Particles
 in $\ee$ Collisions at $\sqrt{s}=130 -209$~GeV
\end{center}
\bigskip
\bigskip
\begin{center}
\LARGE
The OPAL Collaboration \\
\bigskip
\bigskip
\bigskip
\end{center}
\begin{abstract}
A search for stable and long-lived massive particles of electric charge 
$|Q/e|$~=~1 or fractional charges of $2/3$, $4/3$, and $5/3$ is reported
using data collected by the OPAL detector at LEP, at centre-of-mass energies from 130 to 209~GeV.
These particles are assumed to be pair-produced in $\ee$ collisions and 
not to interact strongly.
No evidence for the production of these particles was observed.
Model-independent 
upper limits on the production cross-section between
0.005 and 0.028~pb have been derived for scalar and spin-1/2 particles 
with charge $\pm$1.
Within the framework of the Constrained
Minimal Supersymmetric Standard Model (CMSSM),
this implies a lower limit of 98.0 (98.5) GeV on the mass of 
long-lived right (left)- handed scalar muons and scalar taus. 
Long-lived charged heavy leptons and 
charginos are excluded for masses below 102.0~GeV.
For particles with fractional charge $\pm$2/3, $\pm$4/3 and $\pm$5/3,
the upper limit on the production
cross-section varies between 0.005 and 0.020~pb.
All mass and cross-section limits are derived at the 95\% 
confidence level and are valid for particles with lifetimes
longer than 10$^{-6}$~s.
\end{abstract}
 
\bigskip
\smallskip
\begin{center}
{\large (To be submitted to Physics Letters B)}\\
\bigskip
\bigskip
\bigskip
\end{center}
 

\end{titlepage}
 

\begin{center}{\Large        The OPAL Collaboration
}\end{center}\bigskip
\begin{center}{
G.\thinspace Abbiendi$^{  2}$,
C.\thinspace Ainsley$^{  5}$,
P.F.\thinspace {\AA}kesson$^{  3}$,
G.\thinspace Alexander$^{ 22}$,
J.\thinspace Allison$^{ 16}$,
P.\thinspace Amaral$^{  9}$, 
G.\thinspace Anagnostou$^{  1}$,
K.J.\thinspace Anderson$^{  9}$,
S.\thinspace Arcelli$^{  2}$,
S.\thinspace Asai$^{ 23}$
% ...
}\end{center}\bigskip
\bigskip
$^{  1}$School of Physics and Astronomy, University of Birmingham,
Birmingham B15 2TT, UK
\newline
$^{  2}$Dipartimento di Fisica dell' Universit\`a di Bologna and INFN,
I-40126 Bologna, Italy
\newline
$^{  3}$Physikalisches Institut, Universit\"at Bonn,
D-53115 Bonn, Germany
\newline
% ...
$^{  5}$Cavendish Laboratory, Cambridge CB3 0HE, UK
\newline
$^{  6}$Ottawa-Carleton Institute for Physics,
Department of Physics, Carleton University,
Ottawa, Ontario K1S 5B6, Canada
\newline
% ...
$^{  9}$Enrico Fermi Institute and Department of Physics,
University of Chicago, Chicago IL 60637, USA
\newline
% ...
$^{ 16}$Department of Physics, Schuster Laboratory, The University,
Manchester M13 9PL, UK
\newline
% ...
$^{ 22}$Department of Physics and Astronomy, Tel Aviv University,
Tel Aviv 69978, Israel
\newline
$^{ 23}$International Centre for Elementary Particle Physics and
Department of Physics, University of Tokyo, Tokyo 113-0033, and
Kobe University, Kobe 657-8501, Japan
\newline
% ...

 \newpage
\section{Introduction}

\label{sec:intro}

Many searches for massive new particles predicted by extensions
to the Standard Model (SM) assume that these particles decay
promptly at the primary interaction vertex. 
Such searches are not sensitive to long-lived heavy particles
which do not decay within the detectors. 
There exist, however, a number of models which predict such long-lived 
particles.
For example, in the Constrained Minimal Supersymmetric Standard Model (CMSSM), 
for certain choices of the parameters, sleptons or charginos could be 
long-lived~\cite{ref:MSSM}. 
R-parity violating supersymmetric (SUSY) models~\cite{ref:rpv} also allow for 
long-lived heavy particles and a fourth-generation heavy lepton could be 
stable~\cite{ref:heavy-lept}.
In gauge-mediated supersymmetry, if the SUSY-breaking energy scale is 
sufficiently high~\cite{ref:gmssm}, sleptons could be long-lived.
Some models beyond the SM also predict the existence of
particles with fractional electric charge.
Previous searches for long-lived massive charged particles have been
performed by the LEP collaborations with data taken at the Z$^0$ 
resonance~\cite{ref:zo-stable}, as well as with data taken at higher 
centre-of-mass energies, up to 209~GeV ~\cite{ref:opal-lep2-stable,
ref:lep2-stable}. 

This paper describes an update to a search for long-lived particles, 
referred to here as X$^\pm$, with $m_{\rm X} > m_{\rm Z}/2$,
and charge $|Q/e|$~=~1 or 2/3, pair-produced in the
reaction $\ee \rightarrow {\rm X}^{+} {\rm X}^{-}$.  This search has 
been described in detail in~\cite{ref:opal-lep2-stable}. 
In this paper we also search for particles 
with fractional charges 4/3 and 5/3. All fractionally-charged particles 
are assumed to be colourless and non-strongly-interacting. 
To make the search for these particles as model independent as possible,
only minimal calorimetric information has been used.
Due to their large mass these particles would have anomalously high or low
ionization energy loss $\dedx$ in the tracking chambers.
This search is therefore primarily based on the precise $\dedx$
measurement provided by the OPAL jet chamber. 
The data were collected by the OPAL detector 
during 1995-2000, at centre-of-mass energies 
from 130~GeV to 209~GeV corresponding to a total integrated luminosity 
of 693.1~pb$^{-1}$ as reported in Table~\ref{tab:cands}.

\section{The OPAL Detector}
\label{sec:opaldet}

A complete description of the  OPAL detector can be found 
in Ref.~\cite{ref:OPAL-detector}. Here only a brief overview is given.
The central detector comprised
a system of tracking chambers,
providing track reconstruction
over 96\% of the full solid 
angle\footnote
   {The OPAL right-handed coordinate system is defined such that the $z$-axis is 
    in the
    direction of the electron beam, the $x$-axis points towards the centre of the 
    LEP ring, and $\theta$ and $\phi$ are the polar and azimuthal angles, defined 
    relative to the $+z$- and $+x$-axes, respectively. The radial coordinate is 
    denoted by $r$.}
inside a 0.435~T uniform magnetic field parallel to the beam axis. 
It consisted of a two-layer
silicon microstrip vertex detector, a high-precision vertex drift chamber (CV) with axial and stereo wires,
a large-volume jet chamber and a set of $z$-chambers measuring 
the track coordinates along the beam direction. 

The jet chamber (CJ) is the most important detector for this analysis.
It was divided into 24 azimuthal sectors, each equipped with 159  
sense wires.  Up to 159 position and $\dedx$
measurements per track were thus possible, with a 
precision of $\sigma_{r\phi} \approx 135 \, \mu$m and $\sigma_z \approx $~6~cm.
When a track was matched with $z$-chamber hits and hits on the stereo 
wires of the vertex chamber (CV), the uncertainty
on its $z$ coordinate was $\approx $~1~mm. 
The tracking detectors, located inside the magnet coil, 
provided a track momentum measurement with a resolution of
$\sigma_p/p \approx \sqrt{(0.02)^2 + (0.0015
\cdot p_t)^2}$ for tracks with the full number of hits 
($p_t$, in GeV, is the momentum transverse to the beam
direction) and a resolution on the ionization energy loss measurement 
of approximately $2.8\%$ for $\mumu$ events with a large number of
usable hits for $\dedx$ measurement~\cite{ref:dedx}.

A lead-glass electromagnetic 
calorimeter (ECAL) located outside the magnet coil
covered the full azimuthal range with good hermeticity
in the polar angle range of $|\cos \theta |<0.984$.
The magnet return yoke was instrumented for hadron calorimetry
covering the region $|\cos \theta |<0.99$ and was surrounded by
four layers of muon chambers.
Electromagnetic calorimeters close to the beam axis 
completed the geometrical acceptance down to 24 mrad
on both sides of the interaction point.

\begin{figure}[p]
\centering
\epsfig{file=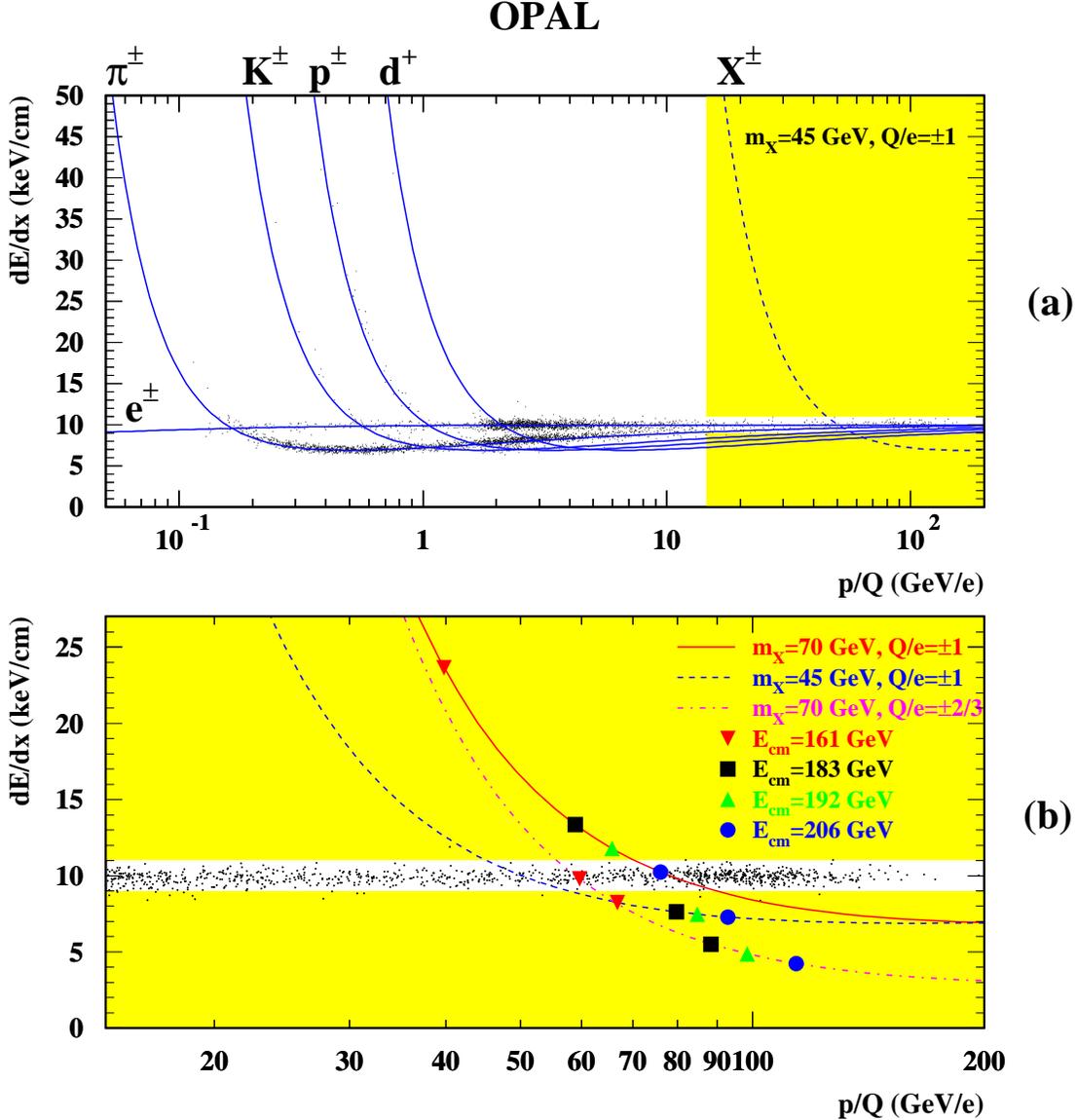,width=15cm}
\caption[]{\sl
  \protect{\parbox[t]{15cm}{
(a) The distribution of the ionization energy loss, $\dedx$, of tracks in the 
CJ detector, as a function of the apparent momentum, $p/Q$, for a sample of the data 
collected in the year 2000.
The two shaded regions are the search regions. The momentum lower limit of the 
search regions is defined by the preselection cut $p>0.07\sqrt{s}$. No cut has 
been applied to the data, apart from a cutoff of $p_t>0.1$~GeV made to 
reject low momentum tracks trapped in the jet chamber volume.\\
(b) Expanded view of the search regions. 
The theoretical curves for heavy long-lived particles are shown with example points
from various centre-of-mass energies.
In $\ee\rightarrow\XX$ 
events, the momentum of the {\rm X}$^\pm$ particles of a given mass 
is fixed by $\sqrt{s}$. 
} } } 
\label{fig:dedx-demo}
\end{figure}

The ionization energy loss $\dedx$ produced by a charged particle 
is a function of the electric charge $Q$ and of $\beta\gamma=p/m$, where $p$ is the momentum and $m$ the mass of the particle~\cite{ref:dedx}.
Figure~\ref{fig:dedx-demo} shows the distribution of $\dedx$ as a function of 
the apparent momentum, $p/Q$. 
Standard particles of charge $\pm$1 (e, $\mu$, $\pi$, p, K) 
with high momentum ($p > 0.1 \sqrt{s}$ ) have $\dedx$ 
between 9 and 11~keV/cm. 
Massive particles with charge $\pm$1 are expected to yield
$\dedx>11$~keV/cm for high-mass values, $m_{\rm X} >0.36 \sqrt{s}$, 
or $\dedx<9$~keV/cm for
low-mass values, $m_{\rm X} < 0.27 \sqrt{s}$. 
The $\dedx$ measurement therefore provides a good tool for particle 
identification in these high- and low-mass regions.
Massive particles with charge $\pm$2/3 would have $\dedx>11$~keV/cm for high mass values, $m_{\rm X} >0.45 \sqrt{s}$ or  $\dedx<9$~keV/cm for low-mass values, $m_{\rm X} <0.35 \sqrt{s}$. The expected $\dedx$ for massive particles of charge $\pm$4/3 and $\pm$5/3 is greater than 11 keV/cm for all mass ranges.  The search for massive particles with charge $\pm$1/3 was not possible because the typical $\dedx$ deposit of these particles would be too close to the instrumental noise level.  

\section{Monte Carlo Simulation}
\label{sec:MC}

Several different Monte Carlo programs were used to generate the signal
process $\ee \rightarrow {\rm X}^{+}{\rm X}^{-}$.
Signal events of the type $\ee \rightarrow \sell^+
\sell^-$ ($\sell^{\pm}$ being a charged scalar lepton)
were generated at several centre-of-mass energies up to 206 GeV
using SUSYGEN~\cite{ref:SUSYGEN}. The generated charged scalar leptons
are not allowed to decay, therefore simulating a signal from 
heavy charged stable scalar particles.
Similarly, events of the type $\ee \rightarrow \LL$ and 
$\ee \rightarrow \QQ$, 
where L$^\pm$ are stable heavy spin-1/2 leptons, and Q,~$\bar{\mathrm Q}$ 
are colourless stable heavy spin-1/2 particles with charge 2/3, 4/3, and 5/3, were
generated at the same energies, using the EXOTIC~\cite{ref:EXOTIC} generator.
All signal samples were generated with 1000 events per mass point 
with mass $m_{\rm X}$ ranging from 45~GeV to the kinematic limit 
for the centre-of-mass energy considered. 
The mass points were generated every 5 GeV with a finer binning of 1 GeV in the
mass regions where we expect lower selection efficiencies.
For the purpose of detector simulation and particle interactions,
all particles were treated as heavy muons. The simulation of $\dedx$ 
in the central jet chamber of OPAL accounted for the charge, mass and 
momentum of the particle, as described above.

The background was estimated using simulations of all 
Standard Model processes (two-fermion, four-fermion and
two-photon processes) for all centre-of-mass energies from 130 to 206~GeV. 
Small differences in the centre-of-mass energies of data and background 
Monte Carlo samples have a negligible effect on the analysis. 

The contribution to the background from two-fermion final states was 
estimated using BHWIDE~\cite{ref:BHWIDE}
for the $\ee$ 
final states and KORALZ~\cite{ref:KORALZ} and KK2f~\cite{ref:KK2F} for the 
$\mumu$ and $\tautau$ states. 
Hadronic two-fermion events, $\qq$,
were simulated using PYTHIA~\cite{ref:JETSET1}. 
For the two-photon background, the PYTHIA~\cite{ref:JETSET1}, 
PHOJET~\cite{ref:PHOJET} and HERWIG~\cite{ref:herwig} Monte Carlo 
generators 
were used for $\ee \qq$ final states
and the Vermaseren~\cite{ref:VERMASEREN} and the BDK~\cite{ref:BDK} generators 
for all $\ee \ell^+ \ell^-$ final states.
Four-fermion final states were 
simulated 
with grc4f~\cite{ref:grace4f}, which takes into 
account interference between all diagrams. 
All generated signal and background events were processed
through the full simulation of the OPAL detector~\cite{ref:GOPAL};
the same event analysis chain was applied to the simulated events
and to the data.

\section{Data Analysis}
\label{sec:analysis}

Pair-produced stable or long-lived massive charged particles would manifest
themselves in events with two back-to-back tracks. 
Assuming they would not interact strongly, these particles would not produce hadronic showers. 
Since they are massive, they would not produce electromagnetic showers either. 
For these reasons, the considered events would be very similar to $\mumu$
events, the only difference being the higher mass of the particles, 
which yields a different $\dedx$ for the same momentum.

A preselection similar to the one described in~\cite{ref:opal-lep2-stable} is used 
for the analyses. Several criteria have been loosened in order to increase the sensitivity to high-mass particles. The criteria are listed below: 

\begin{itemize}
\item[{\bf P1}]
        {Events are rejected if the total multiplicity of tracks in the central
        detector and clusters in the ECAL is greater than 26.
        Cosmic ray events are rejected as in~\cite{ref:leptpairs}. Bhabha events 
        are identified by requiring two energetic and collinear clusters in the 
        electromagnetic calorimeter, these events are then rejected.}
\item[{\bf P2}] 
        {Events are required to contain at least two tracks in the central 
        detector, each satisfying basic quality criteria\footnote
        {The distance between the beam axis and the track at the point of
        closest approach (PCA) must be less than 1~cm; the $z$-coordinate of
        the PCA must be less than 40~cm; the innermost hit of the track 
        measured by the jet chamber must be closer than 75~cm to the beam
        axis.}
        and having a momentum $p>0.07\sqrt{s}$, a momentum transverse to the
        beam axis $p_t>0.025\sqrt{s}$, a polar angle satisfying 
        $|\cos{\theta}|<0.97$ and at least 20 CJ hits usable for $\dedx$ 
        measurement.  The two selected tracks are required to have opposite 
        electric charge.}
\item[{\bf P3}]
        {To reduce background from two-photon events, 
        the total visible energy\footnote{The visible energy, 
        the visible mass and the total transverse momentum of the event
        are calculated using tracks and calorimeter clusters, correcting for 
        double counting as described in~\cite{ref:OPAL-Higgs}.} 
        of the event is required to be $E_{\rm vis}>0.14\sqrt{s}$ and the 
        acoplanarity angle\footnote
        {The acoplanarity angle, $\phi_{\rm acop}$, is defined as 180$\degree$ 
        minus the angle between the two tracks in the $r-\phi$ plane.} 
        between the two tracks is required to be $\phi_{\rm acop}<20\degree$. }
\item[{\bf P4}]
        {Events containing an isolated ECAL cluster with an energy greater than
        5~GeV are rejected to reduce background from events with initial state
        radiation.  Isolation is defined as an angular separation  of
        more than $15\degree$ from the closest track.}
\item[{\bf P5}]
        {It is required that 
        $\frac{E_1}{p_1}+\frac{E_2}{p_2}<0.2$, where $p_{1,2}$ are the momenta of 
        the two selected tracks and $E_{1,2}$
        denote the energies of the ECAL clusters associated to 
        the tracks. This further reduces the contribution from
        Bhabha scattering events.  No other tracks with $p>0.5$~GeV and 
        no unassociated clusters with $E>3$~GeV should be found in a cone of 
        $10\degree$ half-opening angle around each of the two selected tracks}.
        Criterion P5 is not used in the fractionally-charged particles analysis
        since their interaction properties with the calorimeters are unknown.
        This reduces the dependence on calorimeter response around the candidate tracks.
\end{itemize}

After the preselection, the background is dominated by $\ee \rightarrow \mumu$ 
events, with a small contribution from $\ee \rightarrow \tautau$ and 
two-photon $\ee\mumu$ events.  The effect of the preselection cuts on the 
all data samples and various Monte Carlo background processes, for the search for $|Q/e|=$~1 particles, is shown in Table~\ref{tab:cutflow206.6}.

\begin{table}[hbt]
\centering
\begin{tabular}{|c|c||c|c||c|c|}
\hline
 Nominal   &         & \multicolumn{2}{c||}{$|Q/e|=1$ search}           
                     & \multicolumn{2}{c|}{Fractional charge search} \\
\cline{3-6}
$\sqrt{s} (GeV)$ & ${\cal L}$  & candidates & background      & candidates & background    \\
    bins         &(pb$^{-1}$)&            &                 &            &               \\
\hline
  133      & 10.7    &    0       & 0.02$\pm$0.16   &    0       & 0.24$\pm$0.19 \\
  161      & 10.0    &    0       & 0.11$\pm$0.11   &    0       & 0.20$\pm$0.32 \\
  172      & 10.4    &    0       & 0.001$\pm$0.04  &    0       & 0.08$\pm$0.10 \\
  183      & 56.3    &    0       & 0.13$\pm$0.33   &    1       & 0.37$\pm$0.95 \\
  189      & 172.3   &    0       & 0.17$\pm$0.28   &    1       & 0.90$\pm$1.41 \\
  192      & 29.0    &    0       & 0.03$\pm$0.35   &    0       & 0.10$\pm$0.42 \\
  196      & 72.5    &    0       & 0.14$\pm$0.40   &    0       & 0.22$\pm$0.47 \\
  200      & 74.0    &    0       & 0.08$\pm$0.33   &    0       & 0.14$\pm$0.43 \\
  202      & 37.0    &    0       & 0.00$\pm$0.18   &    0       & 0.07$\pm$0.19 \\
  205      & 87.4    &    0       & 0.17$\pm$0.43   &    0       & 0.34$\pm$0.76 \\
  207      & 133.5   &    0       & 0.26$\pm$0.66   &    1       & 0.52$\pm$1.16 \\
\hline
 Total     & 693.1   &    0       &1.1$\pm$1.3  &    3       & 3.2$\pm$2.4\\
\hline
\end{tabular}
\caption[]{\sl
  \protect{\parbox[t]{15cm}{
The number of candidate events and the expected background
at all energies, for the search for $|Q/e|=$~1 and fractionally-charged 
particles.  The errors quoted include both statistical and systematic effects.
In the second column, the integrated luminosity is given for each energy.
The data collected in the year 2000 were delivered at various centre-of-mass energies, up to $\sqrt{s} = 209$~GeV. For this analysis they have been separated in two bins, the first one, referred to as 205~GeV, with $\sqrt{s} < 206$~GeV and an average $\sqrt{s}$ of 204.7~GeV, and the second one, referred to as 207~GeV, with $\sqrt{s} \geq 206$~GeV and an average $\sqrt{s}$ of 206.6~GeV.
} } }
\label{tab:cands}
\end{table}

\subsection{Search for particles with unit charge}
\label{sec:dedx}

\begin{figure}[t]
\centering
\epsfig{file=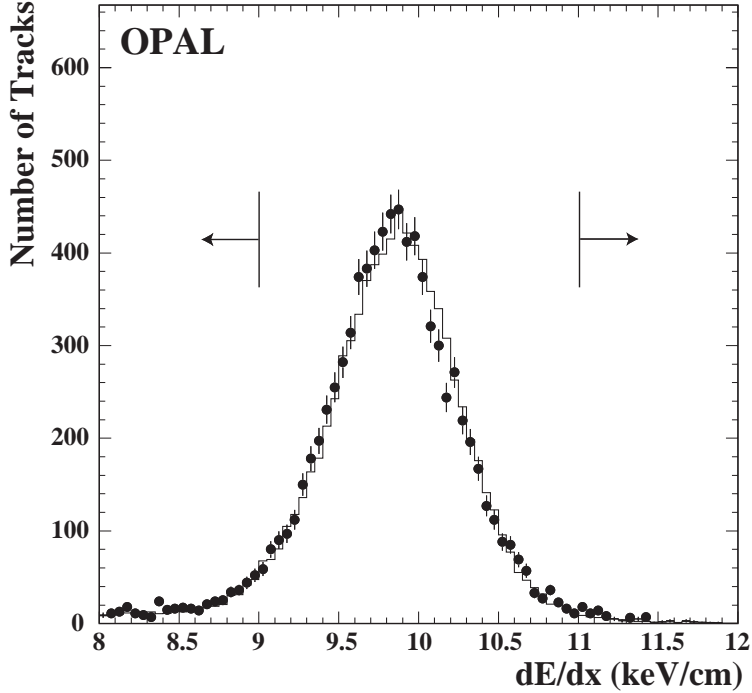,width=10cm} 
\caption[]{\sl
  \protect{\parbox[t]{15cm}{
$\dedx$ distribution for data and Monte Carlo simulation at $\sqrt{s}=189$ GeV after the preselection cut {\bf P4}. The arrows show the accepted region.
} } } 
\label{fig:dedx_dist}
\end{figure}

The search strategy has been simplified with respect 
to~\cite{ref:opal-lep2-stable} and now relies entirely on $\dedx$ information. A sample $\dedx$ distribution for data and simulated Monte Carlo events is shown in Figure~\ref{fig:dedx_dist}.
The kinematic selection is no longer used.
This new strategy has been applied to all data samples. 
The preselected events are retained if they satisfy 
the following requirements on $\dedx$:
\begin{itemize}
\item[{\bf A1}]
{Both high-momentum tracks must have either $\dedx>11$ keV/cm or 
$\dedx<9$ keV/cm. }
\item[{\bf A2}]
{The probability that the $\dedx$ measurements for either track were 
consistent with one of the standard particles (e, $\mu$, $\pi$, p, K) 
must be less than 30\%. This removes background from poorly measured SM particles.}
\end{itemize}

\begin{table}[hbt]
\centering
\begin{tabular}{|l||r||r||r|r|r|r|r||r|r|r|}
\hline
\multicolumn{1}{|c||}{Cuts}      & \multicolumn{1}{c||}{Data}   &
\multicolumn{6}{c||}{Background Simulation} & \multicolumn{3}{c|}{Signal MC (\%)}\\

\cline{3-11}
           &
& \multicolumn{1}{c||}{Total}
& \multicolumn{1}{c|}{$\ee$}
& \multicolumn{1}{c|}{$\mumu$}
& \multicolumn{1}{c|}{$\tautau$}
& \multicolumn{1}{c|}{$\ee\ellell$}
& \multicolumn{1}{c||}{Others}
& $\epsilon_{45}$ & $\epsilon_{55}$ & $\epsilon_{80}$ \\
\hline

 {\bf P1--2}&  20199&  19572.0& 9292.8&   4015.2&   1305.4&   3428.9&  1529.7&96.6&97.5&97.2\\
 {\bf P3--4}&  15935&  15447.4& 7439.8&   3327.2&   1081.0&   2952.8&   646.6&92.5&93.0&94.3\\
 {\bf P5   }&   4995&   4956.6& $<$1.0&   3098.7&    76.5&   1671.3&   110.1&92.3&92.3&94.3\\
\hline
 {\bf A1,A2}&     0&       1.1& $<$1.0&     0.01&    0.01&     1.1&     0.03&85.7&73.4&94.3\\
\hline
\end{tabular}
\caption[]{\sl
  \protect{\parbox[t]{15cm}{
For the $|Q/e|=$~1 analysis, the numbers of events remaining after 
each cut for all data collected at the various centre-of-mass energies 
and for various Monte Carlo
background processes normalised to the
integrated luminosity of the data 
(``{\rm Others}'' refers to $\/ \ee \rightarrow \qq$ 
and $\/ \ee \rightarrow$ four-fermion processes).  
When no candidate events are selected in the Monte Carlo, a 68\% CL upper 
limit on the number of events is used as the statistical uncertainty.
In the last three columns, the efficiencies for $\sell^{+}\sell^{-}$ 
are given (in percent)
for $m_{\rm X}=45, 55, 80$~GeV at $\sqrt{s}=207$~GeV. 
} } }
\label{tab:cutflow206.6}

\end{table}

The effect of cuts {\bf A1-A2} for all data and simulated events
can be seen in Table~\ref{tab:cutflow206.6}.
No candidate event is found. The expected backgrounds at the various centre-of-mass energies are shown in Table~\ref{tab:cands}.  The total background, summed over all energies, is estimated to be 1.1$\pm$1.3 events, where the uncertainty quoted includes both statistical and systematic effects.

The detection efficiency for spin-0 particles 
varies between 75 and 90\% for masses  $m_{\rm X} < 0.27 \sqrt{s}$ or
$m_{\rm X} >0.36 \sqrt{s}$. 
The efficiency drops significantly in the region 
$ 0.27 \sqrt{s} < m_{\rm X} < 0.36 \sqrt{s} $,
but as data sets collected at different
centre-of-mass energies ($\sqrt{s}$) are combined, a reasonable selection efficiency is achieved for all mass values up to close to the kinematic limit. The selection efficiencies of all Monte Carlo samples at all $\sqrt{s}$ are parametrised as a function of $\beta\gamma=p/m_{\rm X}=\sqrt{s/4m_X^2-1}$ of the particle. This parametrisation is used to calculate the efficiency for all masses at each centre-of-mass energy, using linear interpolation.
For spin-1/2 particles, the efficiencies are 2-9\% lower than for 
spin-0 particles due to the 
different angular distribution of the tracks.
We analyse each centre-of-mass energy separately, then combine the results 
for the final cross-section limits.

\subsection{Search for particles with fractional charges}
\label{sec:q23}

To search for particles with fractional charges of 2/3, 4/3, and 5/3 the selection criteria 
{\bf P1} through {\bf P4}, followed by {\bf A1} and {\bf A2} are used. 
The results after each cut for all data samples and various Monte Carlo background processes are shown in Table~\ref{tab:fcutflow206.6}. For charge 2/3 the selection efficiency is between 75 and 90\% for most of the mass range, while for charges 4/3 and 5/3 the efficiency is above 90\% for the whole mass range.
After this selection, one candidate survives in the data sample at 
$\sqrt{s}=183$~GeV, one at $\sqrt{s}=189$~GeV, and one at $\sqrt{s}=207$~GeV, while no candidate is left in any of the other data sets. The masses of the candidates are reconstructed for charge 2/3 using the $\dedx$ and momentum information, while for charge 4/3 and 5/3 kinematic information only is used. The reconstructed masses, track momenta and $\dedx$ values of the candidate events are reported in Table~\ref{tab:cand_info}.
The selected events and the expected background at each centre-of-mass energy are shown in Table~\ref{tab:cands}. 
The total background, summed over all energies, is estimated to be 3.2$\pm$2.4 events, where the uncertainty quoted includes both statistical and systematic effects.

\begin{table}[hbt]
\centering
\begin{tabular}{|l||r||r||r|r|r|r|r||r|r|r|}
\hline
\multicolumn{1}{|c||}{Cuts}      & \multicolumn{1}{c||}{Data}   &
\multicolumn{6}{c||}{Background Simulation} & \multicolumn{3}{c|}{Signal MC (\%)}\\

\cline{3-11}
           &
& \multicolumn{1}{c||}{Total}
& \multicolumn{1}{c|}{$\ee$}
& \multicolumn{1}{c|}{$\mumu$}
& \multicolumn{1}{c|}{$\tautau$}
& \multicolumn{1}{c|}{$\ee\ellell$}
& \multicolumn{1}{c||}{Others}
& $2/3$ & $4/3$ & $5/3$ \\
\hline

 {\bf P1--2}&  20199&  19572.0&  9292.8&   4015.2&   1305.4&   3428.9&   1529.7&93.1&97.6&97.8\\
 {\bf P3--4}&  15935&  15447.4&  7439.8&   3327.2&   1081.0&   2952.8&   646.6&89.7&94.8&94.7 \\
\hline
 {\bf A1,A2}&     3&       3.2&     0.7&     0.03&     0.5&     1.4&      0.5&86.3&94.8&94.7\\
\hline
\end{tabular}
\caption[]{\sl
  \protect{\parbox[t]{15cm}{
For fractional charges analyses, the numbers of events remaining after each cut 
for all data collected at the various centre-of-mass energies
and for various Monte Carlo
background processes normalised to the
integrated luminosity of the data 
(``{\rm Others}'' refers to $\/ \ee \rightarrow \qq$ 
and $\/ \ee \rightarrow$ four-fermion processes).
In the last three columns, the efficiencies
are given (in percent)
for $|Q/e| = 2/3, 4/3, 5/3$ at $ m_{\rm X} = 70$~GeV and $\sqrt{s}=207$~GeV.
} } }
\label{tab:fcutflow206.6}

\end{table}

\begin{table}[hbt]
\centering
\begin{tabular}{|c|c|c|c|c|c|c|c|}
\hline
$\sqrt{s}$  & $p_{1}$ & $p_{2}$ & $(\dedx)_{1}$ & $(\dedx)_{2}$ & \multicolumn{3}{c|} {Masses (GeV)}  \\
\cline{6-8}
     (GeV)    & (GeV)    &   (GeV)   & (keV/cm)  & (keV/cm)  & \small{$|Q/e|=\frac{2}{3}$}      & \small{$|Q/e|=\frac{4}{3}$}    & \small{$|Q/e|=\frac{5}{3}$} \\
\hline
183         &  $64\pm15$ & $19\pm30$ & $11.04\pm0.62$ & $20.48\pm1.00$ & $49.6\pm11.9$ & $95.9\pm9.4$ & -   \\
189         & $30\pm2$ & $29\pm15$ & $11.17\pm0.39$ & $19.85\pm0.95$ & $24.9\pm1.3$ & $85.9\pm1.0$ & $80.7\pm1.6$  \\
207         & $77\pm8$ & $71\pm8$ & $11.06\pm0.34$ & $11.13\pm0.36$ & $61.7\pm4.7$ & - & -  \\
\hline
\end{tabular}
\caption[]{\sl
\protect{\parbox[t]{15cm}{
Information on the candidate events selected by the fractionally-charged analysis. The momentum and $\dedx$ of each track is reported together with the reconstructed masses for the $|Q/e|= 2/3, 4/3, 5/3$ hypothesis. The candidate mass is not reported when the reconstruction procedure gives a kinematically inconsistent result.   
} } }
\label{tab:cand_info}
\end{table}

\subsection{Systematic uncertainties}
\label{sec:syst}

\begin{table}[hbt]
\centering
\begin{tabular}{|c|c|c|c|}
\hline
Quantity & \multicolumn{3}{c|} {Systematic uncertainty (\%)}\\
\cline{2-4}
      & \multicolumn{2}{c|} {Signal} & $\;\;\;$ Background $\;\;\;$\\
\cline{2-3}
      & High efficiency region &
                   Low efficiency region $\;$ & $\;\;\;\;\;\;\;\;\;\;\;\;\;\;\;\;\;\;$   \\
\hline
$\phi_{\rm acop}$   &  \multicolumn{2}{c|} {0.0-1.2} & 0.0   \\
$E_{\rm vis}$     &  \multicolumn{2}{c|} {0.0-2.1} & 0.0 \\
\cline{2-3}
$\dedx$         &  0.0-0.3 & 0.0-27   & $<$0.008 \\
MC statistics   &  0.6-2.6 & 0.8-21   & 58-157 \\
Interpolation   &  0.0-2.7 & 0.0-35   & - \\
\cline{2-4}
MC generator   &  \multicolumn{2}{c|} {-} & 5.9-300 \\
Double tracks  &  \multicolumn{2}{c|} {-} & 2.9  \\
\cline{2-4}
Luminosity      & \multicolumn{3}{c|} {0.22-0.68} \\
\hline
 Total          & 0.7-3.6 &  1.3-39   &  90-309 \\
\hline
\end{tabular}
\caption[]{\sl
\protect{\parbox[t]{15cm}{
Relative systematic uncertainties in the signal efficiency associated with the various quantities used for the $|Q/e| =1$ search.  The ranges given cover all centre-of-mass energies.  The systematic errors vary slightly with centre-of-mass energy, but
strongly with $m_{\rm X}$ for the signal.  For this reason the two regions are reported: high efficiency ($m_{\rm X}/\sqrt{s}<0.27$ or $m_{\rm X}/\sqrt{s}>0.36$) and low efficiency ($0.27 < m_{\rm X}/\sqrt{s} <0.36$).
} } }
\label{tab:syst}
\end{table}

The main sources of systematic uncertainties for this analysis are discussed below and reported in Table~\ref{tab:syst}, for both the signal efficiency and the background estimate:
\begin{itemize}

\item{The errors on the Monte Carlo modeling of $\phi_{\rm acop}$, $E_{\rm vis}$ and $\dedx$ are estimated by comparing the distributions of these variables for data and background Monte Carlo. The relative difference between the averages of the distributions is used to increase or decrease the value of the cut on the relevant variable in order to decrease the overall signal efficiency and background estimate. The difference between the reduced efficiency and the one obtained with the nominal selection is taken as the systematic uncertainty due to the modeling of the variable under consideration. This estimate of the error is more conservative than the one obtained by smearing the $\dedx$ values of the tracks.}

\item{The MC statistical uncertainty, due to the limited number of signal events generated, has been computed using a binomial formula. For background processes the large relative statistical uncertainty is due to the limited number of background events selected.}

\item{The uncertainty due to the linear interpolation of the signal detection efficiency is estimated as the difference between the interpolated values and the efficiency obtained at mass points where MC signal samples were generated, when that mass point was omitted from the interpolation procedure.
}

\item{For background processes of the type $\ee \ell^+ \ell^-$, the Vermaseren~\cite{ref:VERMASEREN} generator has been used as the reference generator. The difference in the background expectation obtained by using BDK~\cite{ref:BDK} instead of Vermaseren has been considered as the uncertainty on the MC generator.}

\item{Two tracks separated by a distance smaller than 2.5 mm could be unresolved and reconstructed as a single track with a high $\dedx$ value.  Events with unresolved double tracks are potential backgrounds for this search.  The systematic uncertainty introduced by the modeling of the double track resolution in the Monte Carlo samples has been estimated to be 2.9 percent.}

\end{itemize}
  
The absolute uncertainty on the background is reported by centre-of-mass energy in
Table~\ref{tab:cands}. The uncertainty introduced on the integrated 
luminosity~\cite{ref:PLB391} is also reported in Table~\ref{tab:syst}.
At a given centre-of-mass energy the different systematic uncertainties are assumed to
be independent, so that the total systematic uncertainty is calculated as the quadratic
sum of the individual uncertainties.
The Monte Carlo modeling of $\dedx$, the Vermaseren-BDK generator, and luminosity 
uncertainties are assumed to be correlated between centre-of-mass energies, while the 
other systematic uncertainties are assumed to be independent.

\section{Results}
\label{sec:results}

The numbers of candidates found in the search for particles 
with charge $|Q/e|$=1 and fractional charges are summarised for all energies in Table~\ref{tab:cands}, together with the expected backgrounds. 
The data show no significant excess above the expected background from Standard Model processes. 

\begin{figure}[h]
\centering
\begin{tabular}{cc}
\epsfig{file=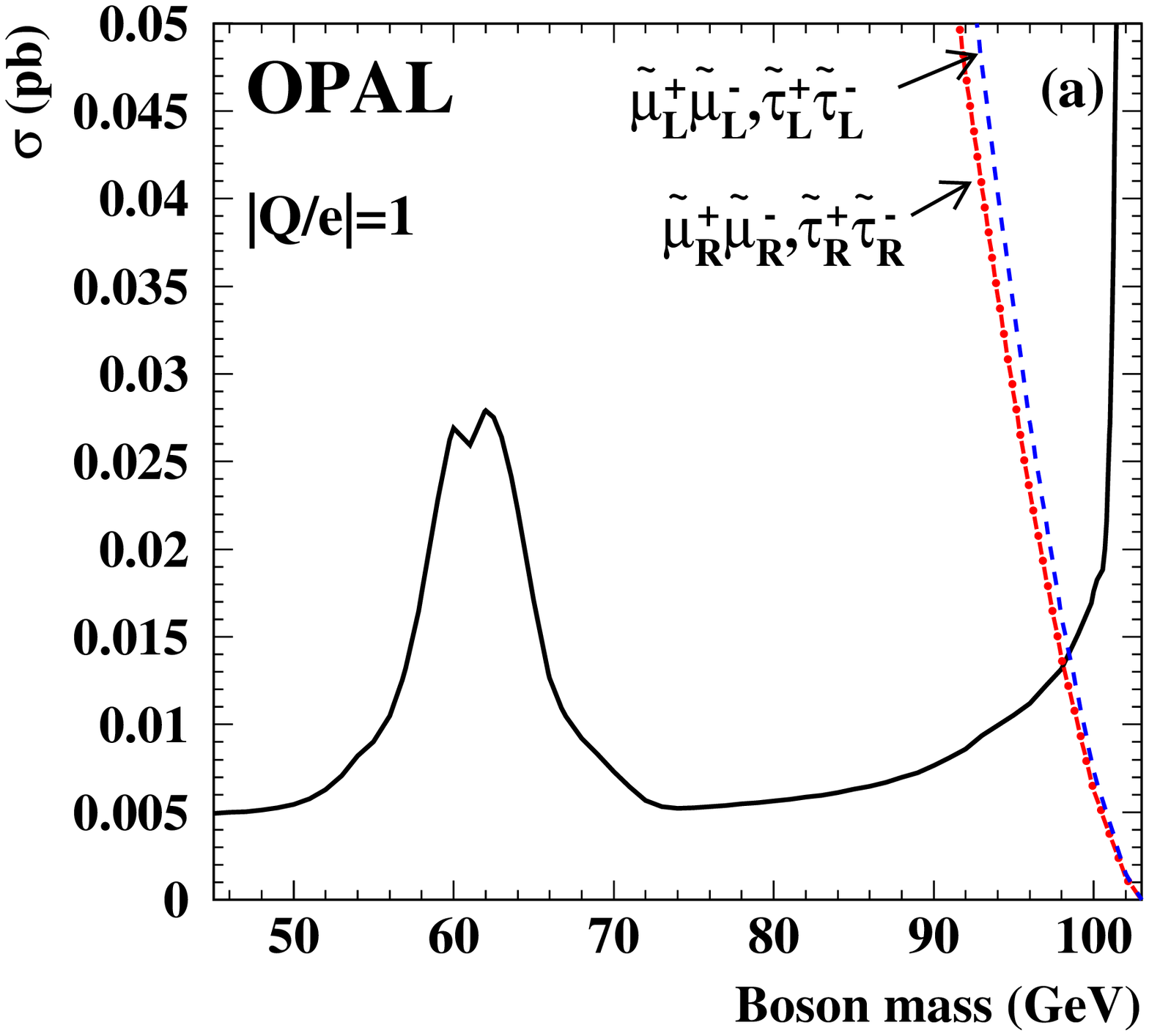, width=8cm} & 
\epsfig{file=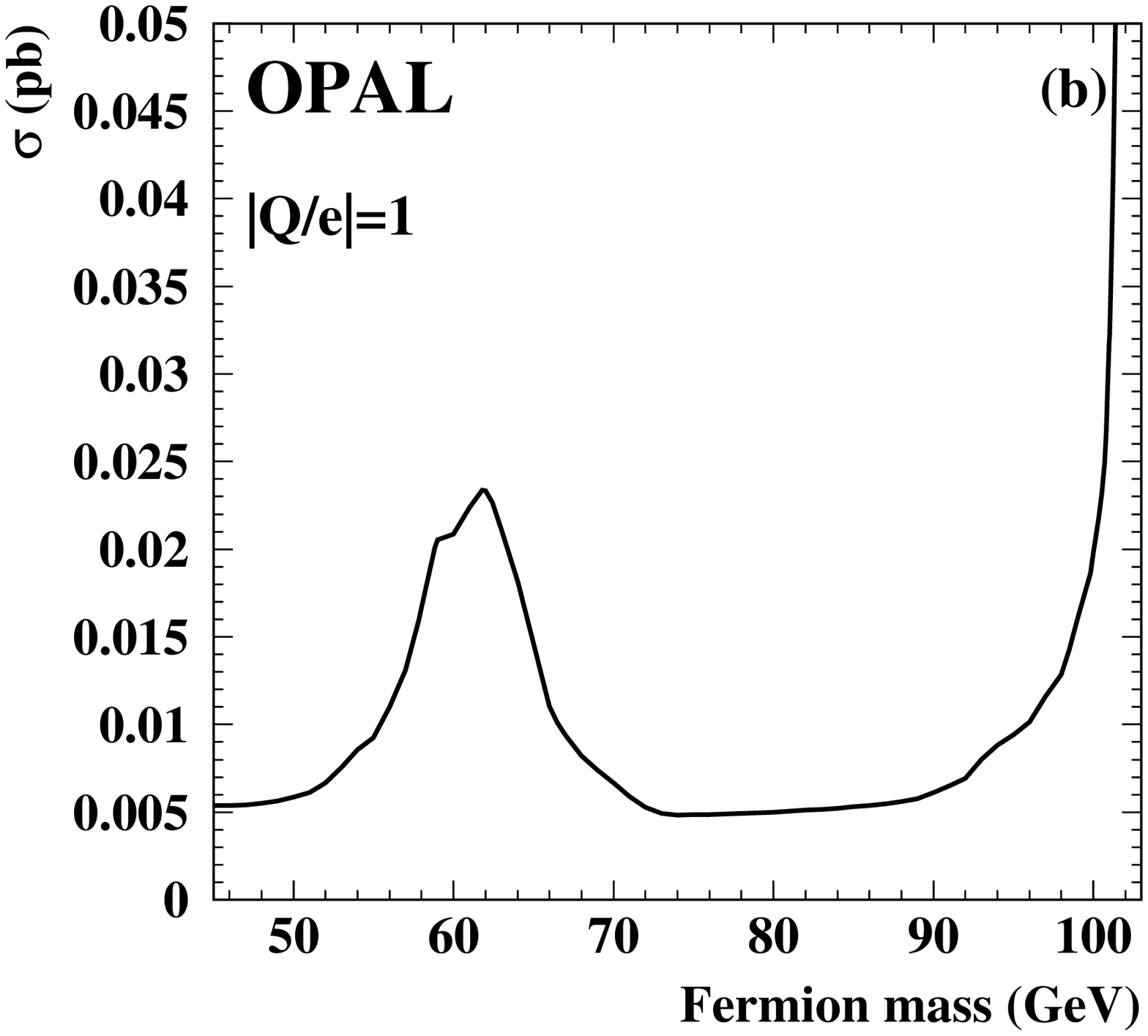, width=8cm}
\end{tabular}
\caption[]{\sl
  \protect{\parbox[t]{15cm}{
Model-independent 95\% C.L. upper limits on the pair-production
cross-section of spin-0 (a) and spin-1/2 (b) heavy long-lived non-strongly-interacting particles of charge $\pm$1 as a function of their mass (solid line) at $\sqrt{s} = 206.6$~GeV. The bump observed between masses of 52 and 70 GeV is due to the drop in efficiency where the $\dedx$ expected for signal crosses the band of $\dedx$ values expected for standard particles. The CMSSM predicted cross-sections for right-handed (dash-dotted line) and left-handed (dashed line) smuons and staus are also shown in (a). 
The 95\% C.L. lower limits on the masses of these sleptons are at the crossing point between the experimental limit and theoretical prediction.
} } }
\label{fig:xsect-scalar}
\end{figure}

Model-independent cross-section upper limits have been computed for the pair-production
of massive charged long-lived particles, combining the results from all centre-of-mass energies, assuming s-channel production.
The cross-section dependence on the energy is taken to be
$\beta^3/s$ for spin-0 particles and $\frac{\beta}{s}(1-\frac{\beta^2}{3})$ 
for spin-1/2 particles, where $\beta=p/E\simeq\sqrt{1-4m_{\rm X}^2/s}$.
In evaluating upper limits, the candidates are counted in mass intervals 
centred on their central mass values and $\pm 2\sigma$ wide (where $\sigma$ is the error on the mass estimate of the candidates as reported in Table~\ref{tab:cand_info}). In the case in which the mass could not be reconstructed the candidates were considered in the whole mass range (from 45 GeV to the kinematic limit for that event).
A likelihood-ratio method~\cite{ref:likelihood} was used to determine an upper limit for the cross-section.
The total systematic error is incorporated into the limits following the prescription of Ref.~\cite{ref:cousins}.

In Figure \ref{fig:xsect-scalar}(a), the 95\% C.L. upper limit 
on the cross-section at $\sqrt{s} = 206.6$~GeV is shown for spin-0 particles
of charge $\pm$~1. 
The 95\% C.L. upper limit on the pair-production cross-section varies 
from 0.005 to 0.028 pb in the mass range $45 < m_{\rm X} < 101$~GeV.  The bump in the mass range of $52 < m_{\rm X} < 70$~GeV is due to the low efficiencies described in section~\ref{sec:dedx}.
The cross-section limits are compared with the predicted cross-sections~\cite{ref:SUSYGEN}
for pair-production of right- and left-handed smuons and staus to determine mass limits. For these two
slepton species, the production cross-section does not depend on the CMSSM
parameters but only on the slepton mass.
The 95\% C.L. lower mass limits are
98.0~GeV and 98.5~GeV for the mass of right- and left-handed smuons and staus, 
respectively, as shown in Figure~\ref{fig:xsect-scalar}(a). 

Figure \ref{fig:xsect-scalar}(b) shows the 95\%~C.L. 
upper limit on the cross-section at $\sqrt{s} = 206.6$~GeV for spin-1/2
particles of charge $\pm$1. The limit varies 
from 0.005 to 0.024 pb in the mass range $45 < m_{\rm X} < 100$~GeV.
This limit must be compared
with the predicted cross-sections for chargino production~\cite{ref:SUSYGEN}
and heavy charged lepton production~\cite{ref:EXOTIC,ref:heavyfermion}.
For the chargino limits, the CMSSM parameters have been chosen to minimise 
the predicted chargino cross-section at every chargino
mass value (assuming a heavy sneutrino, $m_{\sneutrino} > 500$~GeV),
without any restriction on the mass of the lightest neutralino.
A 95\%~C.L. lower limit on
the masses of long-lived charginos, of 102.0~GeV, is obtained for every
choice of the CMSSM parameters.
The 95\%~C.L. lower limit on the heavy charged lepton mass is also 102.0~GeV.

\begin{figure}[!t]
\centering
\begin{tabular}{cc}
\epsfig{file=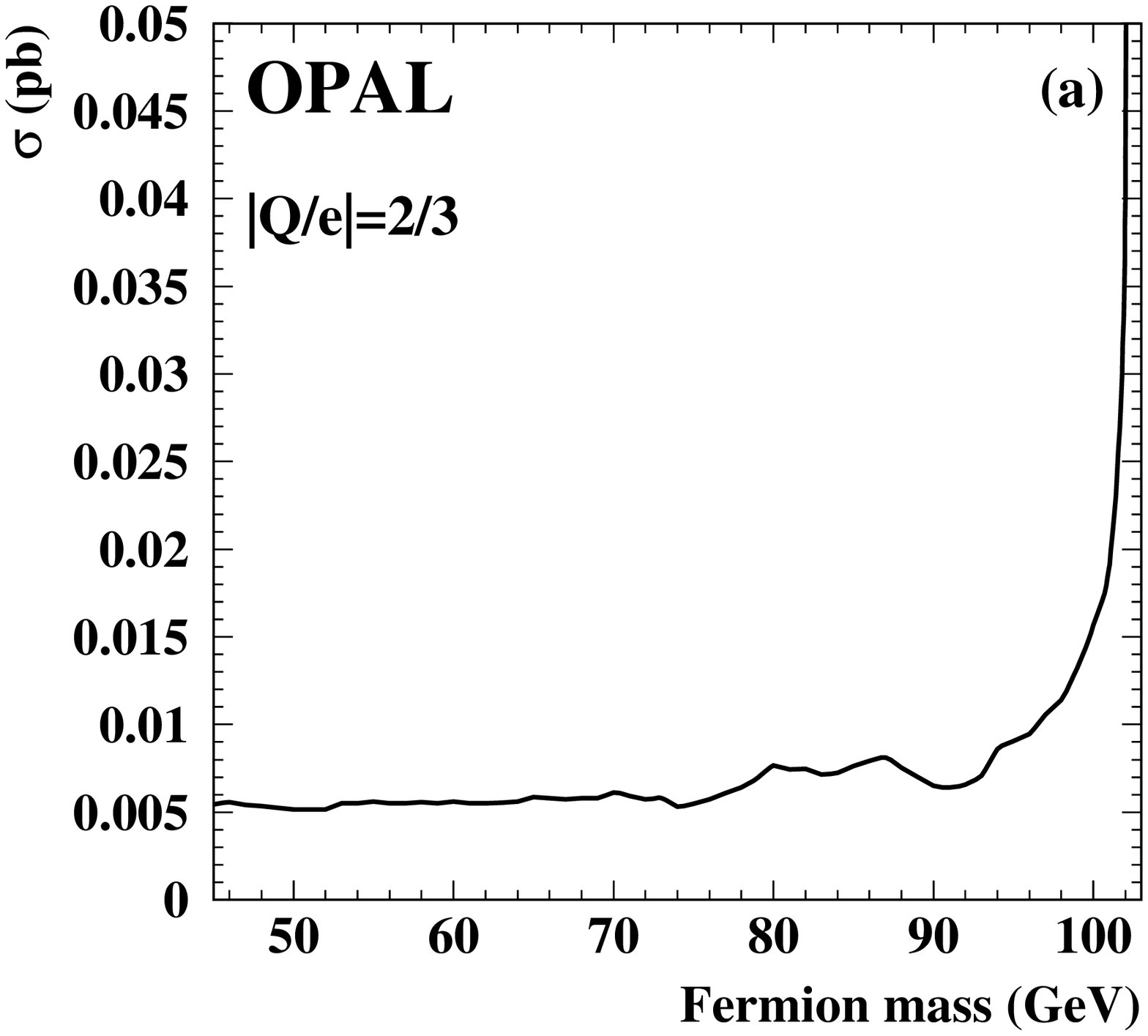,width=8cm} &
\epsfig{file=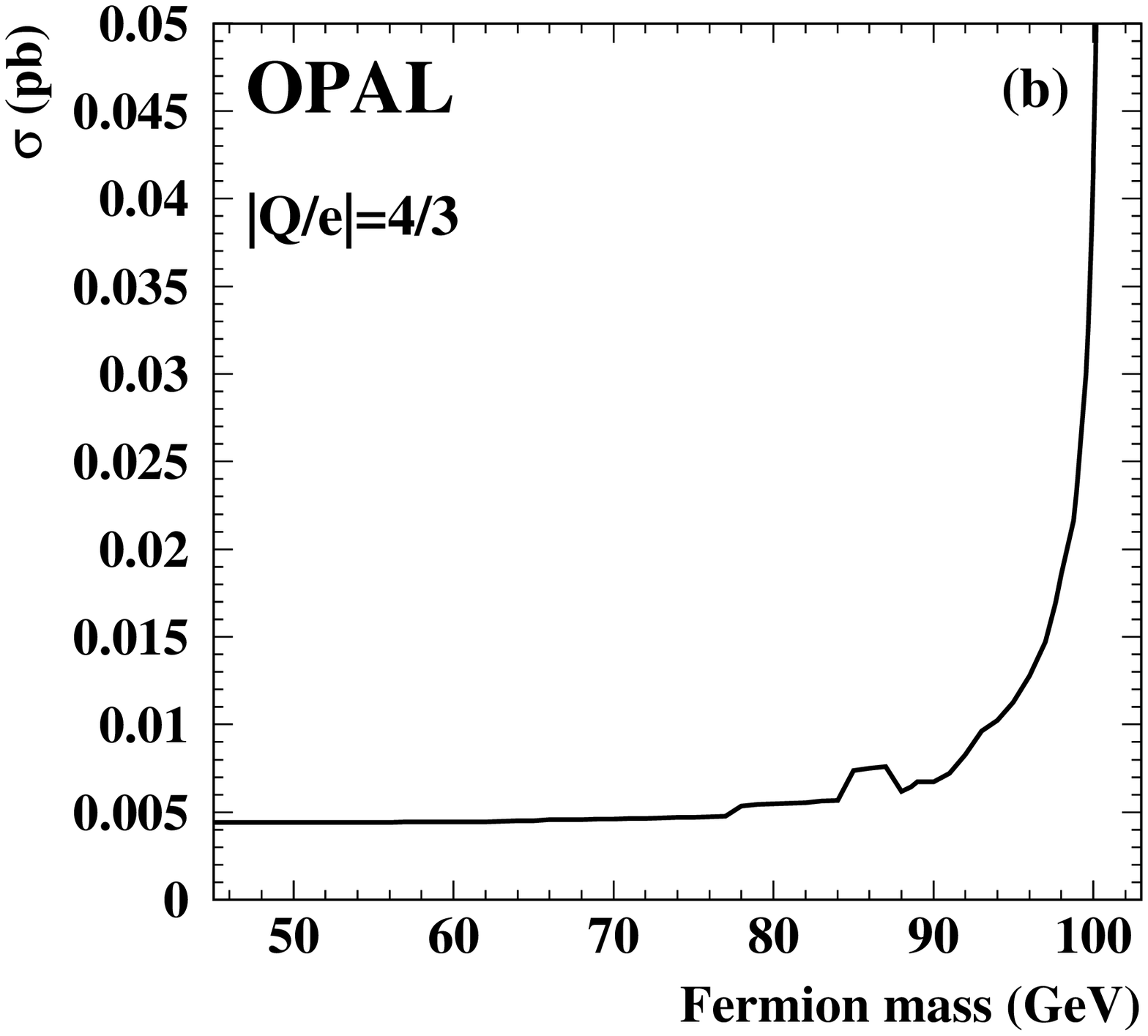,width=8cm} \\
\epsfig{file=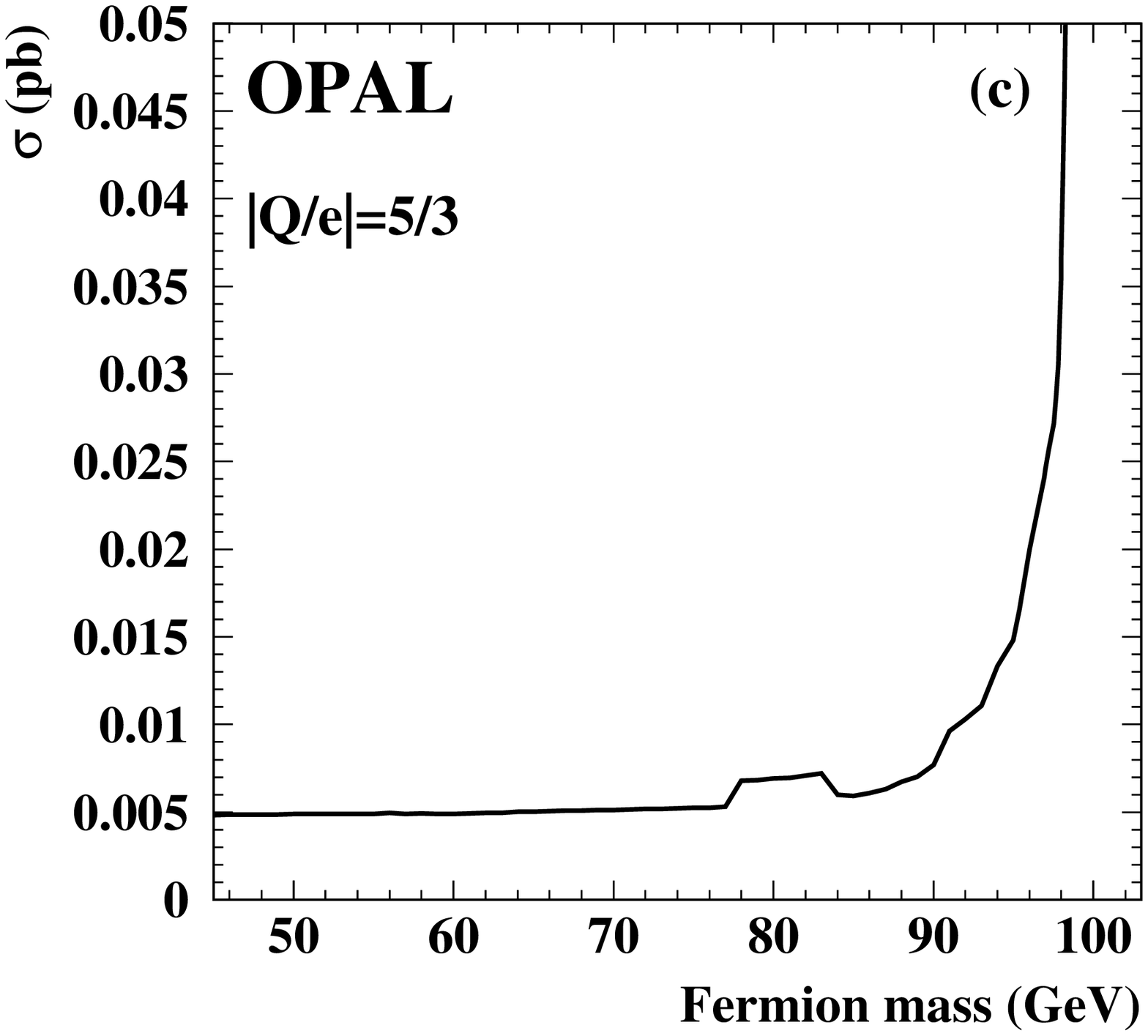,width=8cm} & \\
\end{tabular}
\caption[]{\sl
  \protect{\parbox[t]{15cm}{
Model-independent 95\% C.L. upper limits on the pair-production
cross-section as a function of mass of spin-1/2 heavy long-lived non-strongly-interacting particles of charge (a) $|Q/e|=\pm2/3$, (b) $|Q/e|=\pm4/3$ and (c) $|Q/e|=\pm5/3$, at $\sqrt{s} = 206.6$~GeV.
} } } 
\label{fig:xsect-fermion-q23}
\end{figure}

Figure \ref{fig:xsect-fermion-q23}(a) shows the 95\%~C.L. upper limit on
the cross-section at $\sqrt{s} = 206.6$~GeV for spin-1/2 particles 
of charge $\pm$2/3. The limit varies between 0.005 and 0.020 pb in 
the mass range $45 < m_{\rm X} < 101$~GeV.
Figures \ref{fig:xsect-fermion-q23}(b) and  \ref{fig:xsect-fermion-q23}(c)
show the 95\%~C.L. upper limit on 
the cross-section at $\sqrt{s} = 206.6$~GeV for spin-1/2 particles
of charge $\pm$4/3 and of charge $\pm$5/3, respectively. 
For spin-0 particles with fractional charge the 
cross-section upper limits are slightly higher than for spin-1/2 
particles.  This is due to the difference in angular distributions.

All results obtained are valid for non-strongly-interacting
colourless particles with a lifetime longer than 10$^{-6}$~s.
This lifetime restriction is obtained by considering the heaviest (and 
therefore slowest) particles excluded by this search, and then 
requiring that the decay probability of these particles at a 
flight distance larger than 3.0~m be greater than 95\%. 
For lower mass values the results are also valid for shorter lifetimes.

\section{Summary and Conclusions}
\label{sec:conclusions}

A search was performed for pair-production of stable and 
long-lived massive particles not subject to strong interactions, with 
charge $|Q/e|$~=~1 or fractional charges of $2/3$, $4/3$, and $5/3$.
The primary tool used in this search was the precise $\dedx$ measurement 
provided by the OPAL jet chamber. 
No evidence for the production of such particles
was observed. For s-channel production,
the upper limits on the cross-section 
vary between 0.005 and 0.026~pb in
the mass range explored for particles of charge $\pm$1. 
Within the framework of the CMSSM, lower mass limits have been
obtained: 98.0~GeV for right-handed and 98.5~GeV for the left-handed smuons
and staus. 
Charged long-lived massive leptons and 
long-lived charginos with masses smaller than 102.0~GeV 
are excluded.
For particles with fractional charge $\pm$2/3, $\pm$4/3 and $\pm$5/3, 
the upper limits on the production cross-section vary between 0.005 and 0.020~pb in the range $45 < m_{\rm X} < 95$~GeV.
The above limits are valid at the 95\% C.L. for particles with lifetimes
longer than 10$^{-6}$~s.


\section*{Acknowledgements}

We particularly wish to thank the SL Division for the efficient operation
of the LEP accelerator at all energies
 and for their close cooperation with
our experimental group.  In addition to the support staff at our own
institutions we are pleased to acknowledge the  \\
Department of Energy, USA, \\
National Science Foundation, USA, \\
Particle Physics and Astronomy Research Council, UK, \\
Natural Sciences and Engineering Research Council, Canada, \\
Israel Science Foundation, administered by the Israel
Academy of Science and Humanities, \\
Benoziyo Center for High Energy Physics,\\
Japanese Ministry of Education, Culture, Sports, Science and
Technology (MEXT) and a grant under the MEXT International
Science Research Program,\\
Japanese Society for the Promotion of Science (JSPS),\\
German Israeli Bi-national Science Foundation (GIF), \\
Bundesministerium f\"ur Bildung und Forschung, Germany, \\
National Research Council of Canada, \\
Hungarian Foundation for Scientific Research, OTKA T-038240, 
and T-042864,\\
The NWO/NATO Fund for Scientific Reasearch, the Netherlands.\\




\begin{thebibliography}{99}

\bibitem{ref:MSSM}
H.~P.~Nilles, \PhysRep\ {\bf 110} (1984) 1;\\
H.~E.~Haber and G.~L.~Kane, \PhysRep\ {\bf 117} (1985) 75.

\bibitem{ref:rpv}
H.~Dreiner, {\it ``An Introduction to Explicit R-parity Violation"}, 
Perspective on Supersymmetry, ed. G.L Kane, {\bf 462}, and hep-ph/9707435.


\bibitem{ref:heavy-lept}
E.~Nardi, E.~Roulet and D.~Tommasini, \PhysLett\ {\bf B344} (1995) 225.


\bibitem{ref:gmssm}
S.~Dimopoulos, S.~Thomas and J.~D.~Wells, \NPhys\ {\bf B488} (1997) 39;\\
S.~Ambrosanio, G.~D.~Kribs and S.~P.~Martin, \PhysRev\ {\bf D56} (1997) 1761;\\
G.~F.~Giudice and R.~Rattazzi, \PhysRep\ {\bf 322} (1999) 419.

\bibitem{ref:zo-stable}
\ALEPHColl, D.~Decamp \etal \PhysLett\ {\bf B303} (1993) 198;\\
\DELPHIColl, P.~Abreu \etal \PhysLett\ {\bf B247} (1990) 157;\\
\OPALColl, K.~Ahmet \etal \PhysLett\ {\bf B252} (1990) 290;\\
\OPALColl, R.~Akers \etal \ZPhys\ {\bf C67} (1995) 203.

\bibitem{ref:opal-lep2-stable}
\OPALColl, R.~Ackerstaff \etal \PhysLett\ {\bf B433} (1998) 195.

\bibitem{ref:lep2-stable}
\ALEPHColl, R.~Barate \etal \PhysLett\ {\bf B405} (1997) 379;\\
\DELPHIColl, P.~Abreu \etal \PhysLett\ {\bf B478} (2000) 65;\\
\DELPHIColl, P.~Abreu \etal \PhysLett\ {\bf B503} (2001) 34;\\
L3 Collab., P. Achard \etal \PhysLett\ {\bf B517} (2001) 75.

\bibitem{ref:OPAL-detector}
\OPALColl, K.~Ahmet \etal \NIM\ {\bf A305} (1991) 275;\\
P.~P.~Allport \etal \NIM\ {\bf A324} (1993) 34;\\
P.~P.~Allport \etal \NIM\ {\bf A346} (1994) 476;\\
S.~Anderson \etal \NIM\ {\bf A403} (1998) 326.

\bibitem{ref:dedx}
M.~Hauschild \etal \NIM\ {\bf A314} (1992) 74; \\
M.~Hauschild, \NIM\ {\bf A379} (1996) 436.

\bibitem{ref:SUSYGEN}
S.~Katsanevas and S.~Melachroinos, in {\it ``Physics at LEP2"},
ed. G.~Altarelli, T.~Sj\"{o}strand and
F.~Zwirner, CERN 96-01, vol. 2, (1996) 328;\\
S.~Katsanevas and P. Morzwitz, \CPC\ {\bf 112} (1998) 227.
 
\bibitem{ref:EXOTIC}
R.~Tafirout and G.~Azuelos,
\CPC\ {\bf 126} (2000) 244.

\bibitem{ref:BHWIDE}
S.~Jadach \etal in  {\it ``Physics at LEP2"}, 
eds. G.~Altarelli, T.~Sj\"ostrand and 
F.~Zwirner, CERN 96--01, vol.2,  (1996) 229;\\
S.~Jadach, W.~Placzek and B.F.L.~Ward, 
\PhysLett\ {\bf B390} (1997) 298. 

\bibitem{ref:KORALZ}
S.~Jadach, B.F.L.~Ward and Z.~W\c{a}s, \CPC\ {\bf 79} (1994) 503.

\bibitem{ref:KK2F}
S.~Jadach, B.F.L.~Ward and Z.~W\c{a}s, \CPC\ {\bf 130} (2000) 260.

\bibitem{ref:JETSET1}
T.\ Sj\"{o}strand and M.\ Bengtsson,
          \CPC\ {\bf 43} (1987) 367;\\
{\it ``PYTHIA 5.7 and JETSET 7.4, Physics and Manual"}, CERN-TH.\ 7112/93 (revised August 1995);\\
T.\ Sj\"{o}strand, \CPC\ {\bf 82} (1994) 74;\\
T.\ Sj\"{o}strand, \CPC\ {\bf 135} (2001) 238.

\bibitem{ref:PHOJET}
R.~Engel and J.~Ranft, \PhysRev\ {\bf D54} (1996) 4244;\\
R.~Engel, \ZPhys\ {\bf C66} (1995) 203.

\bibitem{ref:herwig}
G.~Marchesini \etal \CPC\ {\bf 67} (1992) 465.

\bibitem{ref:VERMASEREN}
R.~Bhattacharya, J.~Smith and G.~Grammer, \PhysRev\ {\bf D15} (1977) 3267; \\
J.~Smith, J.A.M.~Vermaseren and G.~Grammer, \PhysRev\ {\bf D15} (1977) 3280; \\
J.~Smith, J.A.M.~Vermaseren and G.~Grammer, \PhysRev\ {\bf D19} (1979) 137.

\bibitem{ref:BDK}
F.A.Berends, P.H. Daverveldt and R. Kleiss, \NPhys\ {\bf B253} (1985) 421;\\
F.A.Berends, P.H. Daverveldt and R. Kleiss, \CPC\ {\bf 40} (1986) 271;\\
F.A.Berends, P.H. Daverveldt and R. Kleiss, \CPC\ {\bf 40} (1986) 285;\\
F.A.Berends, P.H. Daverveldt and R. Kleiss, \CPC\ {\bf 40} (1986) 309.

\bibitem{ref:grace4f} 
J. Fujimoto \etal \CPC\ {\bf 100} (1997) 128.

\bibitem{ref:GOPAL}
J.~Allison \etal \NIM\ {\bf A317} (1992) 47.

\bibitem{ref:leptpairs}
\OPALColl, R. Akers \etal  \ZPhys\ {\bf C61} (1994) 19.

\bibitem{ref:robins}
S.~Robins, {\it ``A Study of Bhabha Scattering at LEP"}, Ph.D. Thesis (1991),
Queen Mary, U. of London, RALT-136.

\bibitem{ref:OPAL-Higgs}
\OPALColl, M.Z.~Akrawy \etal \PhysLett\ {\bf B253} (1991) 511.

\bibitem{ref:PLB391} \OPALColl, K. Ackerstaff \etal \PhysLett\ {\bf B391} 
   (1997) 221.

\bibitem{ref:likelihood}
A.~G.~Frodesen, O.~Skeggestad and T.~Tofte, 
{\it ``Probability and Statistics in Particle Physics"},
Universitetsforlaget, 1979, ISBN 82-00-01-01906-3;\\
S.~L.~Meyer, {\it ``Data Analysis for Scientists and Engineers"}, 
John Wiley and Sons, 1975, ISBN 0-471-59995-6.

\bibitem{ref:cousins}
R.~D.~Cousins and V.~L.~Highland, \NIM\ {\bf A320} (1992) 331.

\bibitem{ref:heavyfermion}
A.~Djouadi, \ZPhys\ {\bf C63} (1994) 317.

\end{thebibliography}
\end{document}